Leveraging Insight from Centuries of Outbreak Preparedness to Improve Modern Planning Efforts

Nina H. Fefferman[1,2], Sharon DeWitte[3], Stephanie S. Johnson[4] and Eric T. Lofgren[4]

1 Department of Mathematics, University of Tennessee, Knoxville, TN
2 Department of Ecology and Evolutionary Biology, University of Tennessee, Knoxville, TN
3 Department of Anthropology, University of South Carolina, Columbia, SC
4 Paul G. Allen School for Global Animal Health, Washington State University, Pullman, WA

Corresponding Author:
Eric T. Lofgren
PO Box 647090
1155 College Ave.
Pullman, WA 99164
Eric.Lofgren@wsu.edu
(509) 335-4022

**Abstract**

Though pandemic preparedness has been a focus of public health planning for centuries, during which our understanding of infectious disease dynamics has grown, our methodologies for managing outbreaks have remained relatively unchanged. We propose leveraging this history to identify opportunities for actual progress. We contrast current plans with historical outbreak control measures and isolate how the complexities of a modern era yield additional challenges in how best to anticipate and mitigate outbreaks. We analyze a diversity of publicly available modern preparedness plans against the context of a historically-based fictional outbreak control strategy described in Defoe's *A Journal of the Plague Year* (published 1720). We identify themes in preparedness planning that remain unchanged from historical settings even though they continue to be actively evaluated in planning efforts. More importantly, we isolate critical modern challenges in preparedness planning that remain predominantly unsolved. These modern, unsolved issues offer best avenues for meaningful improvement. Shifting our planning efforts to focus on identified novel issues may greatly strengthen our local- to global- capacity to deal with infectious threats.

**Introduction**

With the emergence of COVID-19, Ebola, severe acute respiratory syndrome (SARS), pandemic hemagglutinin-1-neuraminidase-1 (H1N1) influenza, as well as the highly publicized emergence of drug-resistant bacteria both in healthcare and within the community, pandemic preparedness has become a topic of not only scientific interest, but of national policy. A great deal of effort has been expended advancing state-of-the-art preparedness, developing sophisticated computational models, developing and improving response plans, and streamlining the development and deployment of pharmaceutical interventions, among others. Many of these preparedness responses, while beneficial, begin building a pandemic response infrastructure from the ground up as the epidemic unfolds. Public health planners can draw on strategies with a long history of success in controlling (though not necessarily eliminating) infectious diseases. Strategies such as vaccination and hygiene/sanitation can be rightly regarded as the foundation from which strides were made in reducing the burden of infectious diseases. Other strategies have less well understood, or more situational benefits, such as quarantine, social distancing and "shelter in place" orders, or the development of social norms that, either by intent or accident, serve to reduce the transmission potential of infectious diseases.

Despite public health being primarily viewed as a relatively modern institution, societies have been coping, or attempting to cope, with sweeping outbreaks of infectious disease since the dawn of recorded history – the Plagues of Athens and Justinian, the recurring waves of bubonic plague that swept Europe, the cholera pandemics of the 1800's and the 1918 influenza pandemic to name but a few. Nor are today's official governmental plans to help mitigate these diseases necessarily new. For example, the word "quarantine" originates from a Venetian policy confining sailors to their ships for a period of time before being allowed into a city(1), and the calls for travel restrictions echo a Florentine attempt to control the plague by banning the importation of certain types of fruit(2) and by prohibition of entry into the city of those who were visibly ill(3). A reflection on the techniques used in the past to attempt to contain outbreaks may yield new insights. It may expose those paths that have yielded the greatest benefits and highlight other weapons in the public heath arsenal that, despite vast leaps in technology, remain fundamentally unchanged. This reflection is especially important during the present COVID-19 pandemic, when a combination of constrained supply chains and no readily available pharmaceutical treatments or vaccines mean, for the moment, that some of the most technologically sophisticated tools used by public health are unavailable.

One of the best described examples of these older preparedness plans comes from Daniel Defoe's *A Journal of the Plague Year* (4), published in 1722. The book is a fictional recounting of a year of life in London during the Great Plague of London in 1665, the last major epidemic of bubonic plague to strike Europe. While not a contemporary account of the plague, though there is some suspicion that it is based on the journal of Defoe's uncle, it is a detailed account of the actual public health practices of the time and the methods used to attempt to slow the spread of the disease (5, 6). By examining the policies recounted in the text, we can consider whether or not current preparedness plans are, in fact, novel, or if there are certain common modes of intervention with roots that pre-date not only modern epidemiology as a field, but also an understanding of Germ Theory.

This understanding, beyond its historical interest, may be beneficial to modern pandemic planning. Planners can identify areas that have not substantially changed in centuries and are not likely to be a source of major innovation as well as identifying "basic" tools that require neither disease specific nor modern administrative support. These tools, in turn, are ones that can be thought of as forming the foundation of a response plan during periods of high uncertainty about the biological and epidemiological properties of the pathogen itself, and where more technologically-driven and pathogen-specific responses may not yet be in place.

A framework can be developed for the design of new plans that work on multiple organizational or administrative levels and can highlight where modern perspectives, capabilities and challenges have resulted in a new environment in which infectious diseases spread. These modern changes may potentially have a negative impact on our ability to protect the public's health that need to be addressed, such as the extent to which disparate locations are connected through transportation or logistical systems. They may also be positive changes which can be exploited, such as vastly improved nutrition and access to medical care, or the development of pharmaceutical interventions such as vaccines or antivirals. Finally, some of the differences between historical and modern preparedness plans may represent a new context that must be addressed, such as changes in governmental scope and scale, and the increasing specialization of healthcare. It is this latter category that represents a relatively untapped resource for the development of new pandemic plans which, to this point, have largely focused on either novel pharmaceutical interventions, or increasingly more detailed and sophisticated versions of public health policies that have been in practice for centuries.

**Methods**

*Comparing Preparedness Plans*

To be able to use the insights from the preparedness plan described in *Journal of the Plague Year*, we first need to correct for medical or scientific understanding which has been improved since the 1700s, translating the concepts into a setting with a modern understanding of

Germ Theory, public health, etc. We must therefore abstract the underlying concepts to capture their intents and consequences rather than focusing on the task-specific recommendations. Once abstracted, we can compare the categories of endeavors recorded and investigate modern preparedness plans to discover whether modern efforts to produce preparedness plans are meaningfully different or whether direct analogies exist between recommendations despite the differences in our modern environment (see Table 1).

In reading both the historical and modern preparedness plans, we discover only 8 individual categories of recommended activities for effective prevention/management/mitigation of disease outbreaks: 1) Designating key players and responsibilities for enforcing policies, 2) Preventative social distancing (targeting either A: those already infectious or B: those remaining susceptible to interrupt potential routes of exposure and transmission), 3) Environmental hygiene, 4) Education, communication, and outreach, 5) Instituting/enforcing boundaries and borders, 6) Detection and surveillance, and 7) Protection of/care for response personnel and 8) Pharmaceutical interventions.

In nearly every case, the modern recommendations are in direct correspondence with historical equivalents, merely updated to reflect a modern understanding of how to accomplish the goals of that category of effect. The only category in which Defoe is silent, and thus the only recommendation unique to the modern era, is that of pharmaceutical intervention; the addition of vaccines, antibiotics, antivirals, etc. greatly strengthens our arsenal in preparing for and containing pandemics.

This direct equivalence between nearly every aspect of historical and modern preparedness planning suggests there is little to be achieved by trying in earnest to discover new potential sets of actions that could improve our safety. Even the widespread shelter in place orders brought into being in the current pandemic are unusual only in the breadth to which they have been applied, and the novel challenges of implementing them primarily questions of political will and economic consequence.

    We therefore strongly suggest that any effort to design new preparedness plans use this framework and, rather than trying to determine from first principles what types of actions might be taken, instead simply consider which categories apply to the case under consideration and how to respond accordingly in that context. This is not at all to recommend that efforts to improve preparedness should cease. Instead, efforts should be based on exploring how improved safety and security can be extracted by leveraging the aspects of the modern era which were functionally inaccessible to earlier societies.

**Results**

*The Unique Challenges in Protecting the Public Health in the Modern Era*

    Naturally, the question that suggests itself is "Which are the relevantly novel aspects of the modern era?" Some of the answers to this question are equally self-evident. Scientific and technological advancement are certainly the most immediate changes in our efforts to prepare for epidemiological threats. Our understand of how to prepare for and/or affect the course of infectious disease outbreaks is enhanced by a number of advances. The emergence of medicine as a scientific field, including the isolation of disease-causing agents is perhaps the foremost, accompanied by improvements in our understanding of immunology, microbiology, epidemiology and physiology, as well as the now-available pharmacopoeia of preventative and therapeutic treatments. Science, however, has not provided a panacean set of solutions such that biosafety and preparedness are solved problems, or ones with inherently technological solutions,

especially early in an emerging epidemic. As our scientific understanding has increased, so has our realization of the scope of diversity and complexity of the threats. While some medical advances provide across-the-board increases in health and resilience to disease challenges, such as improved nutrition or hand hygiene, few provide such broad protection as to be effective against the host of potential pandemic threats we face such that no further effort would be required. Similarly, technological advances in medicine have improved the sensitivity and specificity of diagnostic testing, analysis of data for detection/surveillance, the scope and sophistication of contact tracing and delivery mechanisms for a variety of therapies. In examining how the modern era has changed relative to the past in these regards, nothing is fundamentally changed – preparedness planning must consider the capabilities and limitations of the understanding and methods available for each type of anticipated threat and decide how best to exploit the tools available. The tools have changed, but not their impact.

Another fundamental difference of our modern era is the rapidity of cascading effects across temporal and spatial scales, and across different facets of our daily lives. Many of our societies rely on time-sensitive delivery of goods or provision of services to maintain safe and efficient function. More than any period in the past, a true quarantine that allows nothing in or out of any large-enough area is likely to lead very quickly to a practical crisis more akin to historical sieges than historical public health efforts(7). This has become readily apparent as various state and national-level shutdowns have disrupted international shipping and supply chains during COVID-19, confusion and conflict over the import and export of medical supplies, and competition at the national and even sub-national level for equipment and tools necessary to address the outbreak. Vast swaths of the globe produce insufficient food for their own maintenance, even in the very short term. Water treatment facilities require trained staff support.

Even the removal of municipal solid waste often relies on access to long distance shipping – an acute problem that emerged during the 2014 Ebola outbreak when municipalities refused delivery of potentially infectious biomedical waste.

Beyond the simple basic needs for safe food and water and a sanitary environment, modern risks frequently involve subtle tradeoffs in types of safety and security. Relatively innocuous health risks can lead to widespread panic in ways that ultimately negatively impact large numbers of people (e.g. concern about vaccine safety leading to recent outbreaks of pertussis and measles in the US)(8-11). High economic costs associated with best health practices may compromise the safety and/or stability of a population in the slightly longer term. These costs range from the economic security of specific industries (e.g. cattle diseases and the economic security of agricultural animal farmers (12-14)) to potentially the entire economy in the case of widespread social distancing measures instituted at a state or national level. While these types of tradeoffs have always existed, the rapidity and scale of their interactions and impacts is likely to have truly shifted the ways in which we must weigh their consideration against outbreak preparedness efforts. We clearly should not sacrifice public good in favor of public health; the difficulty lies in figuring out how to balance the two.

Lastly, our modern era differs from earlier contexts in the complexity of many overlapping scales of governance of the facets of our lives and societies. Many of the historical antecedents were based on the declaration of a Mayor or other local official, in London, Florence, or Dubrovnik, local level authorities who could coordinate a single level of response. Modern preparedness often involves plans working on multiple scales: international(15), national (16, 17), state (18-20), and local plans (21, 22), institution-wide plans for schools (23, 24), and healthcare systems (9), and even plans for individual private employers. Critically, this implies that for a plan to be successful, it must survive oversight from and participation by many different individual and organizational parties. These overlapping plans require an understanding of relationships and dependencies between plans, how they are coordinated, and how two or more plans that call for mutually contradictory actions are resolved. This understanding, in turn, requires both research and deliberate planning into how multiscale preparedness systems work when taken as a cascading network of related plans, rather than a collection of distinct entities. Efforts along these lines will not only reduce duplication of effort and help increase the efficacy of integrated responses across scales and nations, but also (perhaps more importantly) by alleviating the need to coordinate in real-time during the critically controllable early stages of an outbreak, we increase our chances of being able to overwhelm potential pandemics before they become global crises (cf. the response to Ebola 2014 (30), in which the outbreak was contained only after coordinated response following an uncertain period during which individual nations spent precious time considering the scale of their contribution to the control efforts(31)).

*Modern Challenges Suggest Modern Opportunities*

Just as these novel elements to modern society increases the complexity of the problem, they also suggest some elements of the process that, by definition, are not as well studied in their potential for improvement in how a preparedness system should plan for or respond to a threat.

Focusing on modern scientific and technological advances is unlikely to yield improvement in preparedness planning. It is impossible to predict the need for specific antibiologicals in advance, and the urgent need for a vaccine does not necessarily correspond to the ease of its development, as evidenced by the long-standing attempts to develop vaccines for Human Immunodeficiency Virus (HIV), dengue fever, or malaria (32-34). Nor does public health urgency necessarily guarantee an availability of funding for development. This suggests that, beyond creating the infrastructure necessary for their development, resources placed toward the development of pharmaceutical interventions in any particular plan cannot be relied upon to bear fruit, at least not in a timely fashion.

Planning for public health response on multiple scales requires a different approach to thinking about the planning problem. Rather than dictating what should be done, which involves enumerating a list of interventions, it shifts the focus to asking who actually makes those decisions, at what time, and what information is needed in order for the decision to be made. The basic toolkits available to public health planners are unlikely to change – with some exceptions, they have not dramatically changed in several centuries – and for many pandemic scenarios, increasingly sophisticated implementations of these same interventions seems unlikely to yield improvements in public health in proportion to the effort involved.

The question of planning on multiple scales, on the other hand, is likely to require detailed, methodical planning. In contrast to intervention-focused plans, it also has the potential to vary widely depending on the scenario being considered. Some scenarios may necessitate "top down" decision making, wherein decisions are made by a centralized authority and then implemented in a distributed fashion. Other scenarios may be better served by a "bottom up" approach, wherein both decision-making and implementation are decentralized and tailored to local needs, and authorities higher up in scale serve primarily to coordinate and enable to sharing of information(35). Similarly, the decisions for some preparedness plans may appropriately lie entirely within the hands of clinicians, while others may require considerable input from policy makers, politicians, or other interest groups. Even which authorities are relevant may change from situation to situation, from a scenario wherein the response is dictated by civilian public health authorities, to ones where the military takes a leading role.

Resolving these questions, over a broad range of potential scenarios, requires the input of a considerable amount of research, political capital, societal investment in maintaining these multiple scale plans over the long term, as well as a public conversation about which levels of authority can and should be responsible for public health decision making. There are well-documented guidelines and trainings available (36,37) as to how to develop incident command systems and emergency operations centers, including the Federal Emergency Management Agency's Emergency Management Institute, but the implementation of these approaches – and their resilience in the face of a true emergency – is highly variable. There are, with near certainty, individuals and organizations who expect to be in a decision making role who will not be, and similarly those anticipating the needed decisions and information will flow from somewhere else when they are, in fact, at the appropriate position and scale to make the decisions themselves. Both are potentially catastrophic misunderstandings of their position within a multi-scale preparedness system. One need look no further than the chaos caused by conflicting federal, state and local-level responses to COVID-19 to see the impact of differing authorities acting on different perceived needs, who may or may not be effectively communicating with one another, nor necessarily providing a clear message to the general public.

The need for an understanding of preparedness plans as multiscale systems, and the current lack of development in this area, suggests a needed transition towards preparedness plans that are not focused on a single siloed set of interventions for a particular disease, each of which are in essence near duplicates of a well-understood set of options. Instead, the development of preparedness plans that emphasize integration of decision making, information providing and implementation on multiple levels, which may both change dramatically based on which scenario is being considered, and represent an aspect of planning that is not necessarily variations on a common and well-understood theme.

**Discussion**

*Better Integration Planning as A Means to Improve Safety is Nothing New*

These recommendations are neither unique to the field of biopreparedness. They are not even novel in the context of emergency management. In 2008, the National Association of County and City Health Officials (NACCHO) conducted an exercise in Montgomery County, Maryland to test a proposed response scenario of quarantine due to an avian flu outbreak. Evaluations of the exercise identified several problems that could affect emergency response, but highlighted how officials relied too heavily on independent information technologies and confusion between state and county roles compromised public health efforts (8). Similarly, during the aftermath of Hurricane Katrina in 2005, emergency officials identified the need for better communication and coordination between emergency personnel as a critical gap in their ability to respond effectively to the situation (9). These are merely two examples among many, but the point remains the same: any endeavor that involves multiple agencies, acting on varieties of scales, and with sometimes conflicting missions will require substantial planning, relationship building between stakeholders, and practice in coordination before all can work effectively and efficiently. There are successful examples of this – the San Diego County Division of Public Health Services conducted a review of its first experience as a designated lead agency during the 2009 novel H1N1 influenza outbreak (40, and a series of training exercises in Minnesota following an outbreak of Highly Pathogenic Avian Influenza helped clarify how industry stakeholders integrated into preparedness efforts (41. Incident command training should and can be done in response to potential threats, not only reviews of emergencies or how to perform for the next time the emergency hits. The United States Department of Agriculture had a four-day tabletop exercise with 14 states, public health labs, and industry to help learn incident command for everyone and potential issues for African Swine Fever (42).

While this is the first time in over a decade the federal incident command system has been tested for a disease, it has a playbook that guides response, policy, and medical countermeasures as well as the minimal agencies that should have representatives and assigned key decision responsibilities depending on risk level (17). It clearly states that 'While States hold significant power and responsibility related to public health response outside of declared Public Health Emergency, the American public will look to the U.S. Government for action when multi-state or other significant public health events occur' showing where strong, clear, concise, and effective messaging and action should start from (17).

**Conclusions**

*Where Do We Go From Here?*

The basic framework of preparedness planning in response to the threat of disease outbreaks has not fundamentally changed in over 300 years. In asking various agencies and actors to generate their own preparedness plans over the past decade, without providing a unified framework by which to do so, we have effectively asked many smart and dedicated people to reinvent the wheel without ever asking if we might instead improve the axle. While it is definitely the case that involvement of crowd-sourced brainstorming from the modern diversity of entities engaged in trying to protect the health of the public and/or continue to function during a public health crisis may have uncovered previously unconsidered elements for inclusion, to the best of our knowledge, no publicly available proposed plan has discovered a fundamentally new category of endeavor. We strongly recommend that, going forward, we refine, adopt, and provide a single, general framework (based on, if not actually the one here presented) as a starting point from which any entity may create their own response plan by effectively filling in the blanks, ***but*** that the efforts required not stop there. Rather than considering the independent response plans from each player to be the endpoint of our preparations, we believe that the completion of these plans should be where the investment in active planning begins. We need to foster, even require by federal, state, and local regulation, the creation of integration/collaboration plans among agencies and actors-both government and industry. We need to dedicate time, energy and thought to research and devise a rubric of best practices for how to integrate action plans among entities who have not explicitly discussed integration of their plans prior to identification of the threat. And resources must be devoted to keeping these plans and their integrations current – in many agencies critical roles may be filled by a single person who may switch jobs or retire, and in the urgency of an ongoing crisis it may be unclear what resources and expertise might be available (e.g. if a local university has experts-or graduate students- available to assist). Too often this sort of response rests on the institutional knowledge of a few key individuals, rather than in a maintainable and transparent system. The current model of presuming that the most critical collaborations will be practiced in advance and all others will be able to act in efficient concert together on the fly is clearly optimistic, at best. We need to define rules of play, consider the modularity of actions, and the transitivity of command. These critical aspects have been predominantly under-studied in favor of parallel efforts to brainstorm individual plans. We believe that, rather than being the afterthought once the plan is finalized, they hold the greatest potential for advancement of the science and social practice of public health preparedness in our modern era.

# References

1. Tognotti E. Lessons from the History of Quarantine, from Plague to Influenza A. *Emerging infectious diseases.* 2013;19(2):254.
2. Palmarocchi R, Compagni D, Villani G. *Cronisti del trecento*. Rizzoli; 1935.
3. Boccaccio G, Steinhöwel H, Leubing H. *Decameron*. Gedruckt auf Kosten des Litterarischen Vereins; 1860.
4. Defoe D. *A journal of the plague year*. Oxford University Press; 2010.
5. Healy M. Defoe's Journal and the English Plague Writing Tradition. *Literature and medicine.* 2003;22(1):25-44.
6. Zimmerman E. HF's Meditations: A Journal of the Plague Year. *Publications of the Modern Language Association of America.* 1972.417-23.
7. Wilkinson A, Leach M. Briefing: Ebola–myths, realities, and structural violence. *African Affairs.* 2014.adu080.
8. Glanz JM, McClure DL, Magid DJ, et al. Parental refusal of pertussis vaccination is associated with an increased risk of pertussis infection in children. *Pediatrics.* 2009;123(6):1446-51.
9. Omer SB, Salmon DA, Orenstein WA, et al. Vaccine refusal, mandatory immunization, and the risks of vaccine-preventable diseases. *New England Journal of Medicine.* 2009;360(19):1981-88.
10. Phadke VK, Bednarczyk RA, Salmon DA, Omer SB. Association Between Vaccine Refusal and Vaccine-Preventable Diseases in the United States: A Review of Measles and Pertussis. *Journal of the American Medical Association*. 2016;315(11):1149-1158.
11. Zucker JM, Rosen JB, Iwamoto M, et al. Consequences of Undervaccination — Measles Outbreak, New York City, 2018–2019. *New England Journal of Medicine*. 2020;382(11):1009-1017.
12. Coffey B, Mintert J, Fox S, et al. The economic impact of BSE on the US beef industry: product value losses, regulatory costs, and consumer reactions. 2011.
13. Pendell DL, Leatherman J, Schroeder TC, et al. The economic impacts of a foot-and-mouth disease outbreak: a regional analysis. *Journal of Agricultural and Applied Economics.* 2007;39(0):19-33.
14. Thompson D, Muriel P, Russell D, et al. Economic costs of the foot and mouth disease outbreak in the United Kingdom in 2001. *Revue scientifique et technique-Office international des epizooties.* 2002;21(3):675-85.
15. *World Health Organization 2019 Novel Coronavirus Strategic Preparedness and Response. 2020.*
16. CDC. *Interim Guidelines for Collecting, Handling, and Testing Clinical Specimens from Persons for coronavirus Disease 2019 (COVID-29). 4.14.2020*
17. Executive Office of the President of the United States. *Playbook for Early Response to High-consequence Emerging Infectious Diseases Threats and Biological Incidents. 2016*18.     Ohio Department of Health. COVID-19 Public Health Orders. 2020.
19. Minnesota Department of Health. Strategies to Slow the Spread of COVID-19 in Minnesota.2020
20. *State of Washington. Washington State Emergency Management Plan, Emergency Support Function 8, Appendix 4-Communicable Disease and Pandemic Response Concept of Operations. 2016.*

21. *D.C. Government Children and Family Services Agency. Pandemic Preparedness plan. 2009.*

22. New York City Department of Health and Mental Hygiene.The New York City Health Care System Preparedness Annual Report . 2017.23. Mt Diablo Unified School District.Coronavirus Response Plan. 2020
24. Hodge J, James G. The legal landscape for school closures in response to pandemic flu or other public health threats. *Biosecurity and Bioterrorism.* 2009;7(1):45-50.
25. Crawford JM, Stallone R, Zhang F, et al. Laboratory surge response to pandemic (H1N1) 2009 outbreak, New York City metropolitan area, USA. *Emerging infectious diseases.* 2010;16(1):8.
26. Schwartz, J., King, C. C., & Yen, M. Y. Protecting Health Care Workers during the COVID-19 Coronavirus Outbreak–Lessons from Taiwan's SARS response. *Clinical Infectious Diseases. 2020.*
27. Bryson-Cahn, C., Duchin, J., Makarewicz, V. A., Kay, M., Rietberg, K., Napolitano, N., et al. A Novel Approach for a Novel Pathogen: using a home assessment team to evaluate patients for 2019 novel coronavirus (SARS-CoV-2). *Clinical Infectious Diseases*.2020.
28. Reusken, C. B., Broberg, E. K., Haagmans, B., Meijer, A., Corman, V. M., Papa, A., et al. Laboratory readiness and response for novel coronavirus (2019-nCoV) in expert laboratories in 30 EU/EEA countries, January 2020. *Eurosurveillance*, 2020.*25*(6).
29. Anthony, C., Thomas, T., Berg, B., Burke, R., & Upperman, J. (2017). Factors associated with preparedness of the US healthcare system to respond to a pediatric surge during an infectious disease pandemic: Is our nation prepared?. *American Journal of Disaster Medicine.* 2017. *12*(4), 203-226
30. World Health Organization. *Ebola Virus Disease outbreak Response Plan in West Africa.* 2014
31. Siedner MJ, Gostin LO, Cranmer HH, et al. Strengthening the Detection of and Early Response to Public Health Emergencies: Lessons from the West African Ebola Epidemic. *PLoS medicine.* 2015;12(3):e1001804-e04.
32. Hsu, D.C. and O'Connell, R.J., Progress in HIV vaccine development. *Human vaccines & immunotherapeutics*, 2017. *13*(5), pp.1018-1030.

33. Vannice KS, Durbin A, Hombach J. Status of vaccine research and development of vaccines for dengue. Vaccine. 2016 Jun 3;34(26):2934-8.
34. Molina-Franky J, Cuy-Chaparro L, Camargo A, Reyes C, Gómez M, Salamanca DR, Patarroyo MA, Patarroyo ME. Plasmodium falciparum pre-erythrocytic stage vaccine development. Malaria Journal. 2020 Dec 19(1):56.
35. *Laboratory Response Network Partners in Preparednes*s.
36. Homeland Security/FEMA. *National Incident Management System. Training Program.* 2011
37. Kohn, S., Barnett, D. J., Galastri, C., Semon, N. L., & Links, J. M. Public Health-Specific National Incident Management System Trainings: Building a System for Preparedness. *Public Health Reports*, *2010. 125*(5_suppl), 43–50.
38. *One Tough Day: Four Communities Prepare for Public Health Emergencies.* 2008.
39. Ball B. Rebuilding Electrical Infrastructure along the Gulf Coast: A Case Study. *BRIDGE-WASHINGTON-NATIONAL ACADEMY OF ENGINEERING-.* 2006;36(1):21.

40. Freedman AM, Mindlin M, Morley C, Griffin M, Wooten W, Miner K. Addressing the gap between public health emergency planning and incident response: Lessons learned from the 2009 H1N1 outbreak in San Diego County. Disaster Health. 2013;1(1):13–20. Published 2013 Jan 1. doi:10.4161/dish.21580

41. Linskens EJ, Neu AE, Walz EJ, et al. Preparing for a Foreign Animal Disease Outbreak Using a Novel Tabletop Exercise. Prehospital and Disaster Medicine. 2018;33(6):640-646. doi:10.1017/S1049023X18000717

42. USDA. *African Swine Fever Exercise.* 2019

**Table 1:** Comparison of Past and Present Response Methods

| CATEGORIES | JOURNAL OF THE PLAGUE YEAR | MODERN RESPONSE PLANS |
|---|---|---|
| **Designating Key Players and Responsibilities for Enforcing Policies** | Explicit designation of roles for existing civil servants *(e.g. Justices of the Peace, Mayors, Bayliffs)*<br><br>Creation of new civil positions in response to the threat *(e.g. Searchers, Watchmen, Keepers, and Buriers)*<br><br>Creation of enforcement duties for both existing and newly created civil servants in overseeing the general public<br><br>Creation of distributed, local supply transport duties to enable sustained quarantine | Explicit designation of roles for existing civil servants *(e.g. state and territorial epidemiologists)*<br><br>Designation of organizational points of contact<br><br>Distribution plans for supplies of medical equipment<br><br>Coordinated reliance on persistent public health-centered agencies<br><br>Creation and Activation of Incident Command Systems and Emergency Operations Centers |
| **Preventative Social Distancing A** *(Isolating Symptomatic People)* | Quarantine of sick people, their households, and possessions<br><br>Altered funeral/burial rites to limit contact with the dead<br><br>Imposed embargo period for use of either public or private modes of transportation after carrying sick people | Self-isolation of the sick<br><br>Physical-barrier equipment (e.g. face masks)<br><br>Encourage healthy behaviors (e.g. cover cough or sneeze, handwashing)<br><br>Encourage fluid absenteeism and sick leave policies at places of employment<br><br>Limit contact with animals |
| **Preventative Social Distancing B** *(Segregating Well People)* | Quarantine of those likely to have been exposed to the sick, or to the property of the sick, but who are not themselves known to be infected | Encourage self-isolation of well individuals<br><br>Encourage short- and long-term travelers (e.g. business travelers and university students living in |

|  |  |  |  |
| --- | --- | --- | --- |
|  | Prohibition against large group activities | dorms, respectively) to return home |  |
|  | Restriction against the sale of second hand items | Employing technology to reduce the need for in-person contact (e.g telework, online classes, telehealth, etc.) |  |
|  | Restricted social contact with others for those handling dead bodies | Encourage healthy behaviors (e.g. cover cough or sneeze, handwashing) |  |
|  | Supplies to quarantined areas should be provided locally, so as not to have suppliers act as routes of transmission | Reduction in large group activities and/or events (e.g. cancelled concerts, sporting events, school classes, etc.) |  |
|  | Imposed curfew on the selling of alcohol to reduce public drunkenness and risks due to inebriation-associated carelessness | Restriction in social gathering places (e.g. closed cafes, theaters, salons, theme parks, malls, etc) |  |
| **Environmental Hygiene** | Locally administered disinfection of any items return to general circulation from areas of quarantine | Increased frequency/reliability of waste removal from public areas |  |
|  | Increased frequency and reliability of street cleaning and waste removal | Increased frequency and care in disinfection of surfaces in public/common areas |  |
|  | Increased distance between occupied residences and garbage dumps | Increased access to disinfectants and cleansers in both public and private spaces |  |
|  | Allocation of responsibility for cleaning of local streets/public property to adjacent landowners | Increased environmental infection control measures in healthcare facilities* |  |
|  | Increased restrictions governing food safety | * this involves a complicated and diverse array of measures, cf. CDC HICPAC |  |
|  | Increased care in funereal/burial sanitation |  |  |

|  |  | Prohibitions against public vagrancy and begging |  |
|  |  | Prohibitions against public presence of domestic or agricultural animals |  |
| **Education, Communication, and Outreach** | Visible marking of quarantined areas<br><br>Visible marking of funeral/burial workers | Public emphasis of appropriate behavioral hygienic measures (e.g. proper hand hygiene, respiratory etiquette, etc.)<br><br>Discouraged participation in large crowds<br><br>Encourage voluntary social distancing<br><br>Encourage fluidity in absenteeism policies in work environments<br><br>Spread awareness of correct information while debunking myths and hoaxes<br><br>Update public on epidemiology of pathogen, and groups most at risk |
| **Instituting/Enforcing Boundaries and Borders** | Restrictions on import/export of goods into the affected region | Institute boundaries for areas providing healthcare to the sick and restrict visitor access and movement of staff to/within the facility<br><br>Limit the number of healthcare personnel entering areas for sick people<br><br>Restrict movement across international borders (in some cases based on health or vaccination status)<br><br>Restrict air travel<br><br>Increase health screening for passengers of air, maritime, and land transportation |

|  |  |  | Increase cargo screening and baggage security |
| --- | --- | --- | --- |
| **Detection and Surveillance** |  | Mandate notification of new case incidence to local authority

Engage public in lay-surveillance reporting

Ensure local responsibility for local detection

Record and report mortality tallies | Establish triage procedures for healthcare workers to separate sick from well people and differentiate causes of illness

Establish testing and reporting protocols and mandates for hospital staff, primary care, and first-response healthcare workers

Send diagnostic samples to regional and/or national laboratories, both private and public

Develop, distribute, and deploy diagnostic tests that balance rapidity of result against accuracy of detection (in outbreak cases, sensitivity may be more important than specificity)

Establish interdepartmental and interagency communications to share testing burdens and surveillance data for analysis (e.g. early warning alert networks, incident command system) |
| **Protection of/Care for Response Personnel** |  | Individual medical staff assigned to care of sick civil servants | Provide prophylactic vaccination to healthcare personnel if available

Increase sensitivity and frequency of disease screening/testing for healthcare personnel

Supply appropriate personal protective equipment, safe protocol training, and pharmaceutical prophylaxis to healthcare personnel

Determine contingency plan for at-risk staff (e.g., pregnant, other defined risk groups) |

|  |  |  | Provide behavioral and mental support to healthcare personnel to mitigate adverse reactions |
|---|---|---|---|
| **Pharmaceutical Interventions** |  | -------------- | Vaccination<br><br>Developing and/or manufacturing and distributing chemical or biological prophylaxis<br><br>Developing and/or manufacturing and distributing chemical or biological targeted infection treatment and/or supportive therapies |